\documentclass[aps,prd,twocolumn,amsmath,superscriptaddress,nofootinbib]{revtex4-1}

\usepackage{varwidth}
\usepackage{multirow}
\usepackage{dcolumn}
\usepackage{tabularx}
\usepackage{booktabs}
\newcolumntype{C}{>{\centering\arraybackslash}X}

\usepackage{graphicx}
\usepackage{epstopdf}
\usepackage{graphics}
\usepackage{graphicx}
\usepackage{epsfig}
\usepackage{epstopdf}
\usepackage{hyperref}
\hypersetup{
    colorlinks=true,
    linkcolor=blue,
    citecolor=blue,
    urlcolor=blue,
    anchorcolor=blue}

\begin{document}
\title{Coalescence production of sexaquark with three diquarks in high-energy nuclear collisions}

\author{Zhi-Lei She}
\affiliation{School of Mathematical and Physical Sciences, Wuhan Textile
            University, Wuhan 430200, China}

\author{An-Ke Lei}
\affiliation{Key Laboratory of Quark and Lepton Physics (MOE) and Institute of
            Particle Physics, Central China Normal University, Wuhan 430079,
            China}
\affiliation{University of Oslo, POB 1048 Blindern, N-0316 Oslo, Norway}

\author{Dai-Mei Zhou}
\email[]{zhoudm@mail.ccnu.edu.cn}
\affiliation{Key Laboratory of Quark and Lepton Physics (MOE) and Institute of
            Particle Physics, Central China Normal University, Wuhan 430079,
            China}

\author{Larissa V. Bravina}
\email[]{larissa.bravina@fys.uio.no}
\affiliation{University of Oslo, POB 1048 Blindern, N-0316 Oslo, Norway}

\author{Evgeny E. Zabrodin}
\affiliation{University of Oslo, POB 1048 Blindern, N-0316 Oslo, Norway}
\affiliation{Skobeltsyn Institute of Nuclear Physics, Moscow State University, Vorob'evy Gory,
Moscow RU-119991, Russia}

\author{Sonia Kabana}
\email[]{Sonja.Kabana@cern.ch}
\affiliation{Instituto de Alta Investigaci$\acute{o}$n, Universidad de Tarapac$\acute{a}$, Arica 1000000, Chile}

\author{Vipul Bairathi}
\affiliation{Instituto de Alta Investigaci$\acute{o}$n, Universidad de Tarapac$\acute{a}$, Arica 1000000, Chile}

\date{\today}

\begin{abstract}
The coalescence production of sexaquark, a hypothetical stable state with quark content $(uuddss)$,
is investigated by the parton and hadron cascade model PACIAE in $pp$ collisions at $\sqrt s = 7$ TeV.
In this work, the compact sexaquark bound state of three diquarks is formed in the final partonic state
by a two-step approach, which involves ``diquark" formation via partonic coalescence and sexaquark construction
with dynamically constrained phase-space coalescence model successively.
The yields, yield ratios, and dependences of spatial parameters (the size of diquark $D_{0}$ and the radius of sexaquark $R_{0}$)
of (anti-)sexaquark are predicted. The yields of a hadronic molecule H-dibaryon $\mathrm{H}(\Lambda\Lambda)$
generated in the final hadronic state are also compared.
These estimates provide references for future sexaquark searches and other exotic state studies, such as dibaryons.
\end{abstract}

\maketitle

\section{INTRODUCTION}
Exotic hadrons with more than the minimal quark content ($q\overline q$ or $qqq$) are allowed
by the quark model proposed by Gell-Mann and Zweig~\cite{gell,zweig}.
In recent years many unconventional hadron candidates containing multiquarks
have been discovered in experiments. As examples, the first observation of
tetraquark candidate $\mathrm X(3872)$ with valence quark content $c\overline c u \overline u$
by Belle Collaboration in 2003~\cite{belle03}, the first observation of
pentaquark state candidates $P_c(4380)$ and $P_c(4450)$ with $c\overline c u u d$
by the LHCb Collaboration in 2015~\cite{lhcb15}. However, their exact nature is still largely unclear.
For a review see Refs.~\cite{mk2018,hx2023,dj2024} and references therein.

Among the multiquark states, the sexaquark (S) with the quark content $uuddss$,
a possible stable multiquark state, has been proposed by G. Farrar~\cite{farrar17}.
Such a state is assumed to have a hard-core radius of
$\sim 0.1-0.4$ fm and mass lower than 2 GeV, and is a good candidate for dark matter
and the composition of neutron stars~\cite{farrar18,farrar22}.
A recent search for sexaquarks has performed and published an upper limit
on its production in $\Upsilon \rightarrow \mathrm{S} \bar \Lambda \bar \Lambda $ decays
by BaBar Collaboration~\cite{babar}.
Note that the sexaquark is different from the H-dibaryon $\mathrm{H}(\Lambda\Lambda)$ with the same quark content,
which was introduced by Jaffe~\cite{jaffe} and estimated to have a mass of 2150 MeV.
In Ref.~\cite{shahrbaf2022} it is suggested that the H-dibaryon
could be a hadronic molecule of two $\Lambda$ hyperons,
while sexaquark is a more tightly compact bound state of three diquarks, bound by color forces.
The hadron resonance gas model (HRGM) with multicomponent hard-core repulsion~\cite{kab18,vvs18}
has been used to study the thermal production rates of sexaquarks
in heavy-ion collisions at LHC energy, where sexaquarks may be produced at relatively high rates~\cite{bdavid21}.

Coalescence mechanism is a useful tool allowing to study exotic multiquark states~\cite{exhic11},
such as the coalescence production of tetraquark $\mathrm X(3872)$ in heavy-ion collisions~\cite{hzhang21,bychen22}.
A dynamically constrained phase-space coalescence model DCPC~\cite{ylyan12}
has been successfully used to describe the production of exotic hadrons with the coalescence of final-state hadrons,
such as X(3872)~\cite{hgxu21,ctwu23}, $P_\mathrm{c}$ states~\cite{chchen22}
in relativistic nuclear collisions, where the hadronic final states were generated
by the parton and hadron cascade model PACIAE 2.0~\cite{paciae2}.
Recently, PACIAE 3.0~\cite{paciae3} + DCPC model has been upgraded to
simulate the exotic hadrons under different internal configurations,
i.e., the compact multiquarks in the final partonic state
and hadronic molecules in the final hadronic state,
such as the X(3872)~\cite{zlshe24} and glueball-like particle X(2370)~\cite{jcao24}.

In the present paper, we employ the newest version of PACIAE 4.0~\cite{paciae4}
and two coalescence models (a partonic coalescence of diquark and the DCPC model)
to estimate the coalescence production of the sexaquark ($\mathrm S$)
with three diquark configuration ($ud-us-ds$) in $pp$ collisions
at the LHC energy assuming such states do exist in nature.
In this study, we expect to investigate the coalescence production
and properties of the exotic (anti-)sexaquark.

The paper is organized as follows: In section~\ref{II}, we briefly describe the PACIAE 4.0
and two coalescence models. Section~\ref{III} contains the simulated results of sexaquark state.
In section~\ref{IV}, a summary is given.

\section{MODELS}\label{II}
As presented above, the PACIAE 4.0~\cite{paciae4} based on the PYTHIA 8.3 code~\cite{pythia8}
and two coalescence models (the partonic coalescence of diquark and the DCPC model) are applied
to simulate the production of the $\mathrm S$ in $pp$ collisions at LHC energy
assuming a coalescence production of the sexaquark.

\subsection{PACIAE MODEL}
The PACIAE model~\cite{paciae2,paciae3,paciae4} is a phenomenological model to simulate
relativistic elementary particle collisions and nuclear collisions.
It divides the whole collision process into four main stages: parton initiation,
parton cascade, hadronization, and hadron cascade.

In PACIAE 4.0 model C\_PY8-simulation framework~\cite{paciae4}, the initial free partons are produced
by turning off hadronization temporarily in PYTHIA 8.3~\cite{pythia8} for each nucleon-nucleon (NN)
or hadron-hadron (hh) collision. Here a special pure quarks-antiquarks initial state is chosen,
i.e., deexcited quarks-antiquarks state.

The parton cascade is further considered by employing the 2 $\to$ 2 parton-parton scattering with
the massive lowest-order (LO) perturbative quantum chromodynamics (pQCD) cross sections~\cite{combridge}.
The resulting final partonic state (FPS) comprises numerous (anti-)quarks
with their four coordinates and four momenta.

After the partonic freeze-out, the hadronization is implemented by
the Lund string fragmentation regime~\cite{pythia6} or the coalescence model~\cite{paciae2}
to generate an intermediate hadronic state.

The hadron rescattering is then performed based on the two-body scattering
until the hadronic freeze-out, resulting in a final hadronic state (FHS) for a NN (hh) collision.

\subsection{COALESCENCE MODELS}
\begin{figure}[htbp]
\includegraphics[scale=0.65]{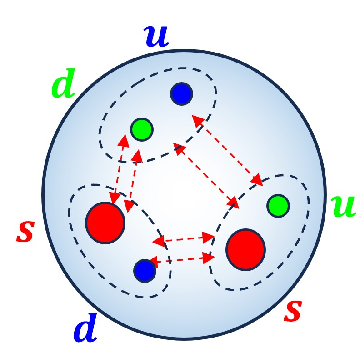}
\caption{The possible structure of the sexaquark as compact bound state of three diquarks ($ud-us-ds$). }
\label{f1}
\end{figure}

A two-step approach is used to produce the exotic sexaquark with three-diquark structure ($ud-us-ds$),
as illustrated in Fig.~\ref{f1}.

First, the ``diquarks" ($uu,ud,dd,us,ds,ss$) are formed via partonic coalescence in FPS,
by matching corresponding $u, d, s$ quark components with the relative distance constraint: $D_{\mathrm {diq}} < D_0$.
Here the $D_{diq}$ is the spatial distance between components in one diquark,
and the $D_0$ is a free parameter of the size of diquark, which can be obtained from theoretical assumptions.
In particular, the value of $D_0$ is 0.4 $\mathrm {fm}$ when the diquark-to-quark $qq/q$ rate
in PYTHIA 8 Monash 2013 tune~\cite{tune2013} is taken into account.

Then these diquarks are further used to coalesce into the sexaquark by DCPC model~\cite{ylyan12}:

In the quantum statistical mechanics~\cite{stowe2007,kubo1965}, the yield of N-particle (such as the partons in FPS) clusters
can be calculated by
\begin{eqnarray}
Y_\mathrm {N}=\int\cdots\int_{E_\alpha\le H\le E_\beta}\frac{d\vec{q}_{1}d\vec{p}_{1}\cdots d\vec{q}_{N}d\vec{p}_{N}}{h^{3N}}
\label{eq: two}.
\end{eqnarray}
where the $E_\alpha$ and $E_\beta$ presents the lower and upper energy thresholds of the particle.
The $\vec{q_{i}}$ and $\vec{p_{i}}$ are the $i$th particle's three coordinates
and three momenta, respectively.
For instance, the yield of $\mathrm S(uuddss)$ assuming to be consisting of three diquarks ($ud$, $us$ and $ds$)
is calculated according to the DCPC model using the following integral:

\begin{eqnarray}
Y_\mathrm {S(uuddss)}=\int \dots \int {\delta_{123}\frac{d\vec{q}_{1}d\vec{p}_{1}d\vec{q}_{2}d\vec{p}_{2}d\vec{q}_{3}d\vec{p}_{3}}{h^9}}
\label{eq: three},
\end{eqnarray}

\begin{eqnarray}
\delta_{123}=\left\{\begin{array}{ll}
1  \mbox { if } 1 \equiv ud,  2 \equiv us,  3 \equiv ds; \\
\quad m_{0}-\Delta m \leq m_\mathrm {inv }\leq m_{0}+\Delta m; \\
\quad  \mathrm {Max}{\{|\vec{q}_{12}|, |\vec{q}_{23}|, |\vec{q}_{13}|\}} \leq R_{0}; \\
0  \mbox { otherwise. }
\end{array}\right.
\label{eq: four}
\end{eqnarray}
\begin{eqnarray}
\begin{aligned}
 m_\mathrm {inv}=\sqrt{\left(E_{1}+E_{2}+E_{3}\right)^{2} -\left(\vec{p}_{1}+\vec{p}_{2}+\vec{p}_{3}\right)^{2}}
\label{eq: five}.
\end{aligned}
\end{eqnarray}

In Eq.~(\ref{eq: four}), $m_0$ represents the rest mass of $\mathrm S$ particle,
and $\Delta m$ indicates the uncertainty of the mass, which can be set
to $m_0=1969.5\,\mathrm{MeV/c^{2}}$ and $\Delta m=84.5\,\mathrm{MeV/c^{2}}$~\cite{shahrbaf2022}.
 $|\vec{q}_{12}|$, $|\vec{q}_{23}|$ and $|\vec{q}_{13}|$ denote the distances
 between each of the three component particles $ud$, $us$, and $ds$
 under the center-of-mass system, respectively,
 while $\mathrm {Max}{\{\dots \}}$ is the maximum distance taken between them.
$R_{0}$ stands for the radius of $\mathrm S$, which can be chosen in the range
of $0.1 \,\mathrm {fm} < R_{0} < 0.4 \,\mathrm {fm}$~\cite{farrar17}.

\section{RESULTS AND DISCUSSION}\label{III}
The main PACIAE 4.0 model parameters of the threshold energy of quark deexcitation and gluon splitting
in the deexcitation process, i.e., the $e_{\rm deex}$ = 2.15 GeV and $e_{\rm split}$ = 0.99 GeV,
are fixed by fitting the ALICE data~\cite{alice201575,alice202181} of
$\pi^{+}+\pi^{-}$, $K^{+}+K^{-}$, $p+\overline p$, and $\Lambda +\overline \Lambda$ yields
in $pp$ collisions at $\sqrt {s}$ = 7 TeV, as shown in Table~\ref{ta1}.
Then the suitable final hadronic state (FHS) and the corresponding final partonic state (FPS)
are obtained with 100 million $pp$ collision events.

After that, the exotic sexaquark ($\mathrm S$) is constructed by the combination
of $ud$, $ds$ and $us$ using the coalescence models (the partonic coalescence of diquark and DCPC model) in FPS.
We use a set of parameters ($0.3 < D_{0} < 0.4\,\mathrm {fm}$,
$0.2 < R_{0} < 0.4\,\mathrm {fm}$, $m_0=1969.5\,\mathrm{MeV/c^{2}}$
and $\Delta m=84.5\,\mathrm{MeV/c^{2}}$) as an illustration in this study.

\begin{table}[htpb]
\caption{The PACIAE 4.0 model simulated yields of $\pi^{+}+\pi^{-}$, $K^{+}+K^{-}$, $p+\overline p$,
and $\Lambda +\overline \Lambda$ are compared to the ALICE data~\cite{alice201575,alice202181}
in $pp$ collisions at $\sqrt{s}=7$ TeV.}	
\centering
\renewcommand{\arraystretch}{1.3}
\begin{ruledtabular}
\begin{tabular}{@{}ccc@{}}
  Particles           & ALICE             & PACIAE  \\ \hline
$\pi^{+}+\pi^{-}$     & $4.49\pm0.20$     & 4.49    \\
$K^{+}+K^{-}$         & $0.572\pm0.032$   & 0.602   \\
$p+\overline p$       & $0.247\pm0.018$   & 0.265   \\
$\Lambda +\overline \Lambda$ & $0.152\pm0.011$ & $0.141$ \\
\end{tabular}
\end{ruledtabular}
\label{ta1}
\end{table}

\begin{figure}[htbp]
\includegraphics[scale=1.0]{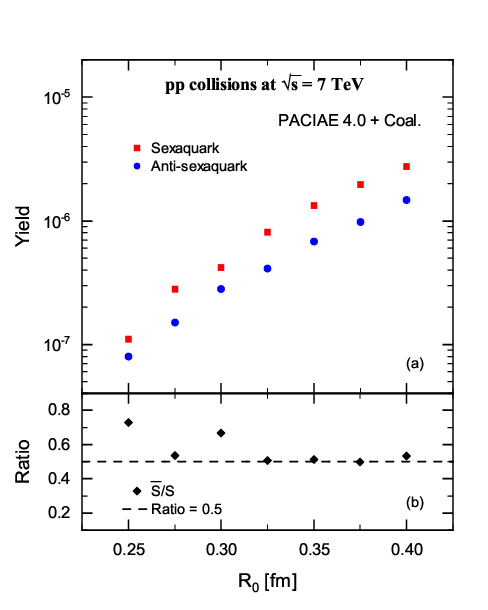}
\caption{The yields of (anti-)sexaquark and the ratios of anti-sexaquark to sexaquark $\mathrm{\overline{S}}/\mathrm{S}$
with the different spatial parameters $R_{0}$ in $pp$ collisions at $\sqrt{s}=7\,\mathrm {TeV}$.
Here the size of diquark $D_{0} = 0.4\,\mathrm {fm}$.
}
\label{f2}
\end{figure}

Figure~\ref{f2} shows the yields of (anti-)sexaquark and the ratios of anti-sexaquark to sexaquark
as a function of spatial parameter $R_0$ in $pp$ collisions at $\sqrt{s}=7\,\mathrm {TeV}$.
Here another spatial parameter $D_{0}$, the size of diquark, is set to $0.4\,\mathrm {fm}$.
In Fig.~\ref{f2}(a), one can see that the yields of (anti-)sexaquark linearly increase with $R_{0}$ on a semi-logarithmic plot,
presenting a significant spatial parameter $R_{0}$ dependence, i.e., $\mathrm{ln} Y \sim R_{0}$.
The values of yields are on the order of $10^{-8}$ to $10^{-6}$ from $R_{0} = 0.25\,\mathrm {fm}$ to $ R_{0} = 0.4\,\mathrm {fm}$.

As shown in Fig.~\ref{f2}(b), the ratio values of anti-sexaquark to sexaquark ($\mathrm{\overline{S}}/\mathrm{S}$)
are close to 0.5, except the fluctuation at smaller $R_{0}$.
The results reflect that the yield of anti-sexaquark is less than that of its corresponding sexaquark
at the same parameter $R_{0}$.
This can be interpreted as anti-particles production is harder than
that of corresponding particles in high energy nuclear collisions~\cite{zls22}.

\begin{figure}[htbp]
\includegraphics[scale=1.0]{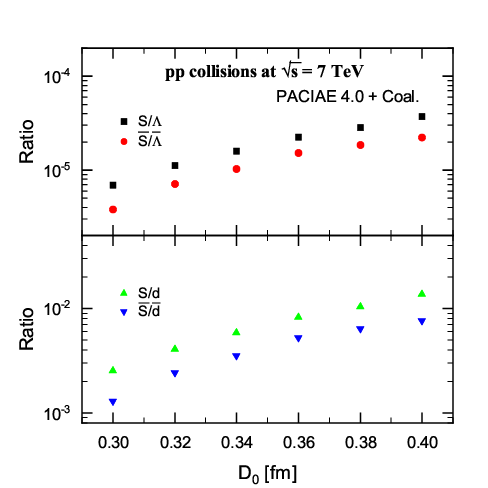}
\caption{The yield ratios of sexaquark ($\mathrm S$) to hyperon ($\Lambda$) and
sexaquark to deuteron ($\mathrm d$) with the different spatial
parameters $D_{0}$ in $pp$ collisions at $\sqrt{s}=7\,\mathrm {TeV}$.
Here the radius of sexaquark $R_{0}$ is in the range of $0.2 < R_{0} < 0.4\,\mathrm {fm}$.
}
\label{f3}
\end{figure}

The yield ratios of sexaquark to hyperon, and sexaquark to deuteron with different spatial parameters $D_{0}$
in $pp$ collisions at $\sqrt{s}=7\,\mathrm{TeV}$ are predicted in Fig.~\ref{f3} and Table~\ref{ta2}
(Only the cases of $D_{0} =0.3\,\mathrm {fm}$ and $D_{0} = 0.4\,\mathrm {fm}$ are shown as examples).
Here the yields of (anti-)sexaquark are the cumulative values in the range of $0.2 < R_{0} < 0.4\,\mathrm {fm}$.
In Fig.~\ref{f3}, the values of yield ratios increase with increasing $D_{0}$,
reflecting that the ratios of coalescence sexaquarks to strange hadrons and light nuclei depend on
the spatial parameters $D_{0}$ of the size of diquark.
More specifically, the ratios of $\mathrm{S}/\Lambda (\mathrm{\overline S}/\overline \Lambda$)
and $\mathrm{S}/\mathrm{d} (\mathrm{\overline S}/\mathrm{\overline d}$) are respectively
on the order of $10^{-6}$ and $10^{-3}$ in $D_{0}=0.3\,\mathrm {fm}$,
while $10^{-5}$ and $10^{-2}$ in $D_{0}=0.4\,\mathrm {fm}$, as Table~\ref{ta2} shows.

Furthermore, we find that the simulated yield ratios are much lower
than that of thermal production in Pb-Pb collisions, e.g.,
the thermal sexaquark with a mass of $1950\,\mathrm{MeV}$ is predicted to have a
rate of $25\%$ of the deuteron in Pb-Pb collisions at a center of mass energy of $2.76\,\mathrm{TeV}$
at the LHC~\cite{bdavid21}. The difference in the predictions between coalescence model
and thermal production may be caused by spatial structure constrains,
selections of thermal conditions, and different collision systems.

\begin{table}[htpb]
\caption{The coalescence yield ratios of sexaquark ($\mathrm S$) to hyperon ($\Lambda$),
and sexaquark to deuteron ($\mathrm d$) in $pp$ collisions at $\sqrt{s}=7\,\mathrm {TeV}$.
Here $D_{0}$ is the spatial parameter of the size of diquark,
and the radius of sexaquark $R_{0}$ is in the range of $0.2 < R_{0} < 0.4\,\mathrm {fm}$.}	

\centering
\renewcommand{\arraystretch}{1.3}
\begin{ruledtabular}
\begin{tabular}{@{}ccc@{}}
 \multirow{2}{*}{Types} &\multicolumn{2}{c}{Ratios} \\ \cline{2-3}
                        &     $D_0=0.30\,\mathrm {fm}$     &   $D_0=0.40\,\mathrm {fm}$  \\ \hline
 $\mathrm{S}/\Lambda $   & $6.91\times 10^{-6}$  & $3.74\times 10^{-5}$        \\
 $\mathrm{\overline S}/\overline \Lambda$   & $3.77\times 10^{-6}$  & $2.22\times 10^{-5}$        \\
 $\mathrm{S}/\mathrm{d}$~\footnote{$\mathrm{d}$ is taken from Ref.~\cite{nar20}.} & $ 2.52 \times 10^{-3}$      & $1.35 \times 10^{-2}$ \\
  $\mathrm{\overline S}/\mathrm{\overline d}$~\footnote{$\mathrm{\overline d}$ is taken from Ref.~\cite{nar20}.} & $1.30\times 10^{-3}$      & $0.74\times 10^{-2}$ \\
\end{tabular}
\end{ruledtabular}
\label{ta2}
\end{table}

To compare with two possible exotic hadrons with the same quark content $(uuddss)$,
i.e., sexaquark $\mathrm{S}(uuddss)$ and H-dibaryon $\mathrm{H}(\Lambda\Lambda)$,
their coalescence yields and mixed ratios $\mathrm{\overline{S}}/\mathrm{S}$,
$\mathrm{\overline{H}}/\mathrm{H}$, and $\mathrm{S}/\mathrm{H}(\mathrm{\overline{S}}/\mathrm{\overline{H}})$
in $pp$ collisions at $\sqrt{s}=7$ TeV are listed in Table~\ref{ta3}.
Here the yield of sexaquark is evaluated with the spatial parameters ($D_{0} = 0.4\,\mathrm {fm}$ and $0.2 < R_{0} < 0.4\,\mathrm {fm}$).
The H-dibaryon is recombined by the components of two $\Lambda$ hyperons
using DCPC model in FHS. The mass and radius of H-dibaryon as hadronic molecule,
if it could exist, are assumed: $2055 \leq  m_\mathrm{inv}(\mathrm{H}) \leq 2230 \,\mathrm{MeV/c^{2}}$~\cite{jaffe},
$1.0 < R_{0} < 4.9\,\mathrm {fm}$~\cite{yka22}, respectively.

\begin{table}[htpb]
\caption{The simulated yields of sexaquark with three diquarks and H-dibaryon with two $\Lambda(uds)$ hyperons,
and the mixed ratios $\mathrm{\overline{S}}/\mathrm{S}$, $\mathrm{\overline{H}}/\mathrm{H}$,
and $\mathrm{S}/\mathrm{H}$ in $pp$ collisions at $\sqrt{s}=7$ TeV.
Here $D_{0} = 0.4\,\mathrm {fm}$ and $0.2 < R_{0} < 0.4\,\mathrm {fm}$ for sexaquark,
and $1.0 < R_{0} < 4.9\,\mathrm {fm}$ for H-dibaryon.}	
\centering
\renewcommand{\arraystretch}{1.3}
\begin{ruledtabular}
\begin{tabular}{@{}ccc@{}}

\multirow{4}{*}{Yields} & $\mathrm{S}(uuddss)$                         &  $2.76\times 10^{-6}$   \\
                        & $\mathrm{\overline{S}}(\overline{uuddss})$   &  $1.47\times 10^{-6} $   \\
                        & $\mathrm{H}(\Lambda\Lambda)$                 &  $ 1.65\times 10^{-4}$   \\
                        & $\mathrm{\overline{H}}(\overline{\Lambda\Lambda})$ &  $ 2.99\times 10^{-5}$ \\  \hline
\multirow{4}{*}{Mixed ratios} & $\mathrm{\overline{S}}/\mathrm{S}$       & $0.533$  \\
                        & $\mathrm{\overline{H}}/\mathrm{H}$             & $0.181$ \\
                        & $\mathrm{S}/\mathrm{H}$                        & $1.67\times 10^{-2}$  \\
                        & $\mathrm{\overline{S}}/\mathrm{\overline{H}}$  & $4.92\times 10^{-2}$ \\
\end{tabular}
\end{ruledtabular}
\label{ta3}
\end{table}

Table~\ref{ta3} shows that the coalescence yields of $\mathrm{S}(\mathrm{\overline{S}})$ are much less than
that of $\mathrm{H}(\mathrm{\overline{H}})$ under the selected parameters.
The mixed ratios $\mathrm{S}/\mathrm{H}(\mathrm{\overline{S}}/\mathrm{\overline{H}})$
indicate that the yield of $\mathrm{S}(\mathrm{\overline{S}})$
is lower than $\mathrm{H}(\mathrm{\overline{H}})$ by two orders of magnitude.
It may be related not only to the spatial structures, formation time, etc.,
but also to the fact that the three-body combinations are more difficult
to generate than two-body combinations in the coalescence framework.
However, the value of $\mathrm{\overline S}/\mathrm{S}$ ($\sim$0.533) is larger than
that of $\mathrm{\overline H}/\mathrm {H}$) ($\sim$0.181) simulated by PACIAE+DCPC model in $pp$ collisions.
This phenomenon suggests that, compared to their corresponding particle counterparts,
the compact anti-H-dibaryon is harder to produce than that of anti-sexaquark.
Further studies are needed to fully understand these yield features.

\section{SUMMARY AND OUTLOOK}\label{IV}
We have employed PACIAE 4.0 and two coalescence models to simulate the sexaquark production
in $pp$ collisions at $\sqrt{s}= 7\,\mathrm{TeV}$.
In this simulation the sexaquark is assumed to be a compact bound state
of three diquark configuration ($ud-ds-us$).
The ``diquark" formation via partonic coalescence and then sexaquark construction
with DCPC model are executed successively in the final partonic state.

The yields of (anti-)sexaquark then are computed with different spatial parameters $R_{0}$
of the radius of sexaquark selected.
The results show a significant spatial parameter $R_{0}$ dependence.
The values are on the order of $10^{-8}$ to $10^{-6}$ from $R_{0} = 0.25\,\mathrm {fm}$ to $ R_{0} = 0.4\,\mathrm {fm}$.
Moreover, the yield of anti-sexaquark is lower than its corresponding sexaquark,
i.e., the ratio values of anti-sexaquark to sexaquark are close to 0.5.

For the yield ratios, the coalescence sexaquarks to hyperons and sexaquarks to light nuclei
present a spatial parameter $D_{0}$ of the size of diquark dependence.
The values are respectively on the order of $10^{-6}$ and $10^{-3}$ in $D_{0}=0.3\,\mathrm {fm}$ case,
while $10^{-5}$ and $10^{-2}$ in $D_{0}=0.4\,\mathrm {fm}$ case, which are much lower
than that of thermal production in heavy-ion collisions.

Besides, although sexaquark $\mathrm{S}(uuddss)$ and H-dibaryon $\mathrm{H}(\Lambda\Lambda)$ have the same quark content $(uuddss)$,
the coalescence yields of $\mathrm{S}(\mathrm{\overline{S}})$ as compact bound state
are much less than that of $\mathrm{H}(\mathrm{\overline{H}})$ as hadronic molecule recombined in final hadronic state,
i.e., their yields differ by two orders of magnitude.
However, the yield ratio of $\mathrm{\overline S}/\mathrm{S}$ ($\sim$0.533) is larger than
that of $\mathrm{\overline H}/\mathrm {H}$ ($\sim$0.181).

The study should be extended to Pb-Pb collisions at LHC energies.
The use of this compact three-diquark configuration for the dibaryons
(such as $p\Lambda$, $\Lambda\Lambda$, and $\Omega\Omega$) also deserves to be studied further.

\section*{Acknowledgements}
The authors thank B.-H. Sa, Y.-L. Yan, W.-C. Zhang, and H. Zheng for helpful discussions.
This work of D.M. Zhou is supported by the National Natural Science Foundation of
China under grant No. 12375135. A.K. Lei and Z.L. She acknowledge
the financial support from the China Scholarship Council.
S. Kabana acknowledges partial support from the Agencia Nacional de
Investigaci$\mathrm{\acute{o}}$n y Desarrollo de Chile (ANID), Chile,
with the grants ANID PIA/APOYO AFB230003, Chile,
and ANID FONDECYT regular No. 1230987, Chile.

\end{document}